\documentclass[twocolumn,amsmath,nofootinbib,amssymb]{revtex4}

\usepackage{epsfig}
\usepackage{graphicx}
\usepackage{dcolumn}
\usepackage{bm}
\usepackage{color}

\def\thebiblio#1{
\begin{center}\bf \large References
\end{center}
\list
{[\arabic{enumi}]}{\settowidth\labelwidth{#1.}\leftmargin\labelwidth
 \advance\leftmargin\labelsep
 \usecounter{enumi}}
 \def\newblock{\hskip .11em plus .33em minus -.07em}
 \sloppy
 \sfcode`\.=1000\relax}

\begin{document}
\preprint{}
\title{%
Dark matter origin of the gamma ray emission \\from the galactic center observed by HESS }

\author{J. A. R. Cembranos, 
V. Gammaldi 
and A.\,L.\,Maroto,
}

\affiliation{Departamento de  F\'{\i}sica Te\'orica I, Universidad Complutense de Madrid, E-28040 Madrid, Spain}%
\date{\today}

\begin{abstract}
We show that the gamma ray spectrum observed with the HESS array of Cherenkov telescopes
coming from the Galactic Center (GC) region and identified with the source
HESS J1745-290, is well fitted by the secondary photons coming from dark matter (DM)
annihilation over a diffuse power-law background. The amount of photons and morphology of
the signal localized within a region of few parsecs,  require compressed
DM profiles as those resulting from baryonic contraction, which offer $\sim 10^3$ enhancements
in the signal over DM alone simulations. The fitted background from HESS data
is consistent with recent Fermi-LAT observations of the same region.
\end{abstract}
\maketitle

Observations of very high energy (VHE) $\gamma$-rays coming from the Galactic Center (GC)
have been reported by different collaborations such as CANGAROO \cite{CANG}, VERITAS \cite{VER},
HESS  \cite{Aha, HESS}, MAGIC \cite{MAG} and Fermi-LAT  \cite{Vitale, ferm}.
In this work, we will focus on the data collected by the HESS collaboration during the years 2004, 2005, and
2006 associated with the HESS J1745-290 source \cite{HESS}. The absence of variability in the TeV data suggests that
the emission mechanism and emission regions differ
from those invoked in the variable IR and X-ray emission \cite{X}. Important deviations from a power law spectrum has
been already proved, and a cut-off at several tens of TeVs is a remarkable feature in the data.
The angular distribution of the VHE $\gamma$-ray emission of HESS J1745-290 shows the presence of an adjunctive diffuse
$\gamma$-ray emission component, but the significance of the signal reduces to few tenths of degree in any case \cite{HESS}.

The fundamental nature of this source is still unclear. These gamma rays could have been originated by particle propagation
\cite{ferm,SgrA} in the neighborhood of the Sgr A East supernova remnant and the supermassive black hole Sgr A, both located
at the central region of our galaxy \cite{Atoyan, AN}. In this letter, we will analyze the possibility of
explaining the spectral features of the signal with the photons produced by the annihilation of DM particles.
This interpretation has been widely discussed in the literature from the very early days of the publication
of the observed data by the above collaborations \cite{Bergstrom1,DMint}. It was concluded that
the spectral features of the data disfavored the DM origin \cite{DMint}. However, in our study, we will assume that
the DM signal is not the only contribution, but it is complemented by a background which is well motivated by radiative
processes generated by  particle acceleration in the vicinity of Sgr A East supernova and the supermassive black hole:
\begin{equation}
\frac{d\Phi_{\text{Tot}}}{dE}=\frac{d\Phi_{\text{Bg}}}{dE}+\frac{d\Phi_{\text{DM}}}{dE}\,.
\label{gen}
\end{equation}
In order to simplify the parameters in our fits, we will assume just a simple power-law for the contribution
not associated with DM:
\begin{eqnarray}
\label{powerlaw}
\frac{d\Phi_{\text{Bg}}}{dE}=B^2 \cdot \left(\frac{E}{\mbox{GeV}}\right)^{-\Gamma}\;.
\end{eqnarray}
This assumption is also experimentally motivated by Fermi-LAT data corresponding to  25 months of observations of  the source
IFGL J1745.6-2900, that is claimed to coincide spatially with the HESS J1745-290 source \cite{Cohen}.
These data have been shown to be well fitted by a broken power-law, that for $E \gtrsim 2$ GeV is consistent
with the spectral index $\Gamma=2.68\pm 0.05$  ($\chi^2/dof=0.81$) \cite{ferm}. In any case, we will not assume any prior on
the parameters for the background, and either $B$ and $\Gamma$ will be fitted from HESS data in our analysis.

The differential gamma-ray flux coming from DM particles can be written in general as:
\begin{equation}
\frac{d \Phi_{\text{DM}}}{dE} =\sum^2_{a=1} \sum^{\text{channels}}_i \frac{\zeta^{(a)}_i}{a} \cdot \frac{dN^{(a)}_{i}}{dE}
 \cdot \frac{{\Delta\Omega\,\langle J_{(a)} \rangle}_{\Delta\Omega}}{4 \pi M^a}\,,
\label{eq:totalflux}
\end{equation}
where $a=2$ takes into account the gamma rays coming from DM annihilation (we are assuming that the DM particle is its
own antiparticle), whereas $a=1$ accounts for the photons generated by the possible DM decays. To determine the Galactic
signal, it is necessary to compute the astrophysical factors $\langle J_{(a)} \rangle$ in the  direction given by the
$\Psi$ angle, defined as the one between the direction of the Galactic center and the line of observation, given by:
\begin{eqnarray}
\langle J_{(a)} \rangle= \frac{1}{\Delta\Omega}\int_{\Delta\Omega}\text{d}\Omega\int_0^{l_{max}(\Psi)} \rho^a [r(l)] dl(\Psi)\,,
\label{flux}
\end{eqnarray}
where $l$ is the distance from the Sun to any point in the halo. The radial distance $r$ is measured from the GC,
and is related to $l$ by $r^2 = l^2 + D_\odot^2 -2D_\odot l \cos \Psi$, where $D_\odot \simeq 8.5$ kpc is the distance from the Sun
to the center of the Galaxy. The distance from the Sun to the edge of the halo in the direction $\theta$ is
$l_{max} = D_\odot \cos \theta + \sqrt{r^2-D_\odot^2 \sin \theta}$. The astrophysical factor is proportional to $\rho^2$ when
accounting for DM annihilation channels (whereas it is just proportional to $\rho$ when computing photons from DM decays).

The photon flux is maximized in the direction of the GC, and must be averaged over the solid angle of the detector.
For detectors with sensitivities in the TeV regime, the solid angles are typically of order
$\Delta \Omega = 2 \pi ( 1 - \cos \Psi ) \simeq 10^{-5}$, as it is the case for the HESS Cherenkov telescopes array.
The dark halo is usually modeled by the NFW profile \cite{Navarro:1996gj}, that is in good agreement with non-baryonic
cold DM simulations.  However, it has been claimed \cite{Blumenthal, Prada:2004pi} that when baryonic gas is taken
into account, it falls to the central region, modifying the gravitational potential and increasing the DM density in the center
(see however \cite{Romano}). This fact has two important consequences for
our analysis. On the one hand, the central region accessible to gamma ray detection is compressed to few tenths
of a degree; on the other hand, the DM annihilating fluxes are enhanced by  a factor $\sim 10^3$ over the classical NFW
profile \cite{Prada:2004pi}. The interpretation of the HESS data from DM annihilation is in good agreement with these types
of compressed dark halos.

\begin{figure}[t]
\begin{center}
\epsfxsize=13cm
\resizebox{7.8cm}{5.6cm}
{\includegraphics{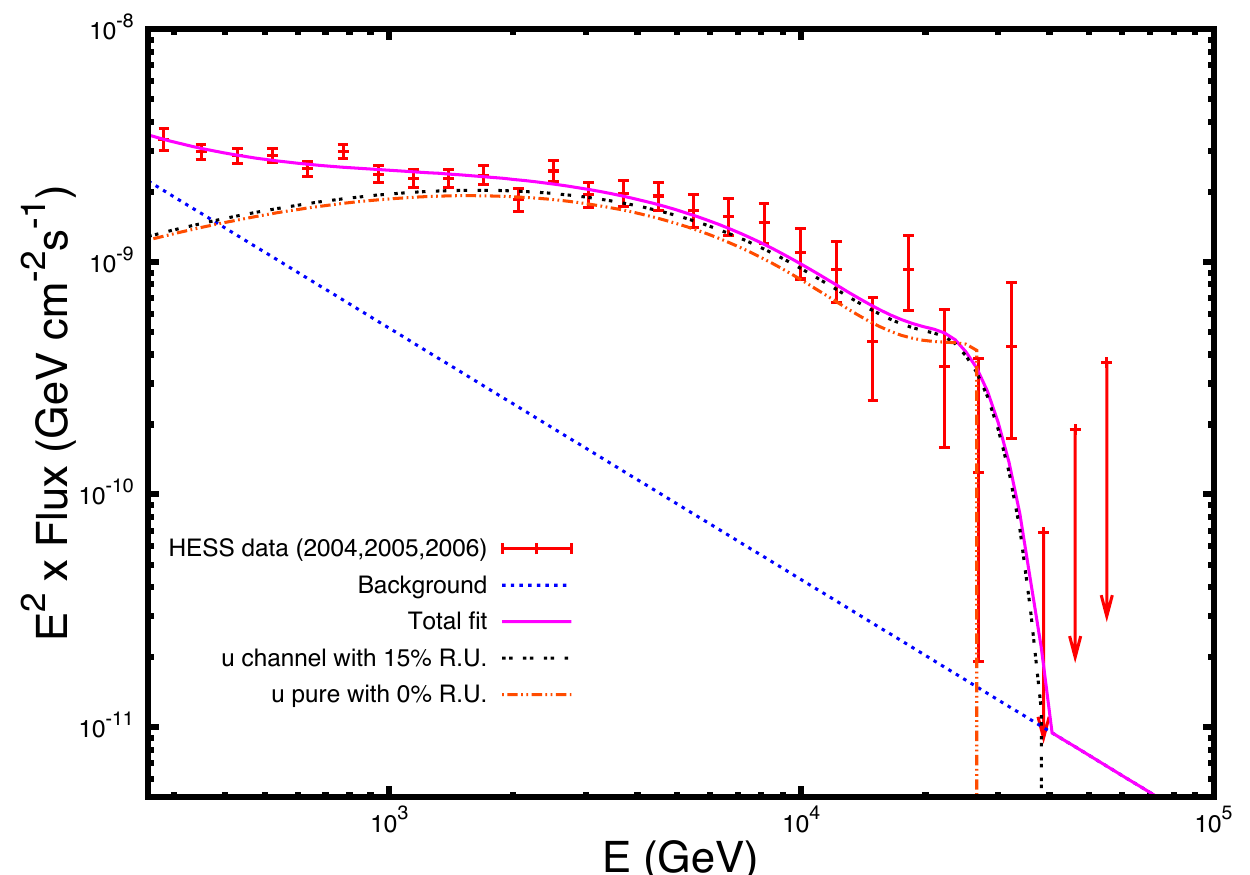}}
\caption {\footnotesize{Best fit to the HESS J1745-290 collection of data (years 2004,2005, and 2006 \cite{HESS}) in the case that the DM
contribution came entirely from annihilation  into $u\bar u$ quarks. The full line shows the total fitting function (Eq. (\ref{gen})).
The dotted line is the fitted power-law background given by Eq. (\ref{powerlaw}). The dot-dashed line corresponds to the DM
annihilation contribution with resolution uncertainty (R.U.) of $15\%$, typical from HESS (the DM contribution
without R.U. is shown by the dotted line for reference). It is remarkable to note that the cut-off in the spectrum characteristic
of this annihilation channel coincides with data. The parameters of the fit are reported on Table \ref{channels}.}}
\label{u fit}
\end{center}
\end{figure}

\begin{figure}[t]
\begin{center}
\epsfxsize=13cm
\resizebox{7.8cm}{5.6cm}
{\includegraphics{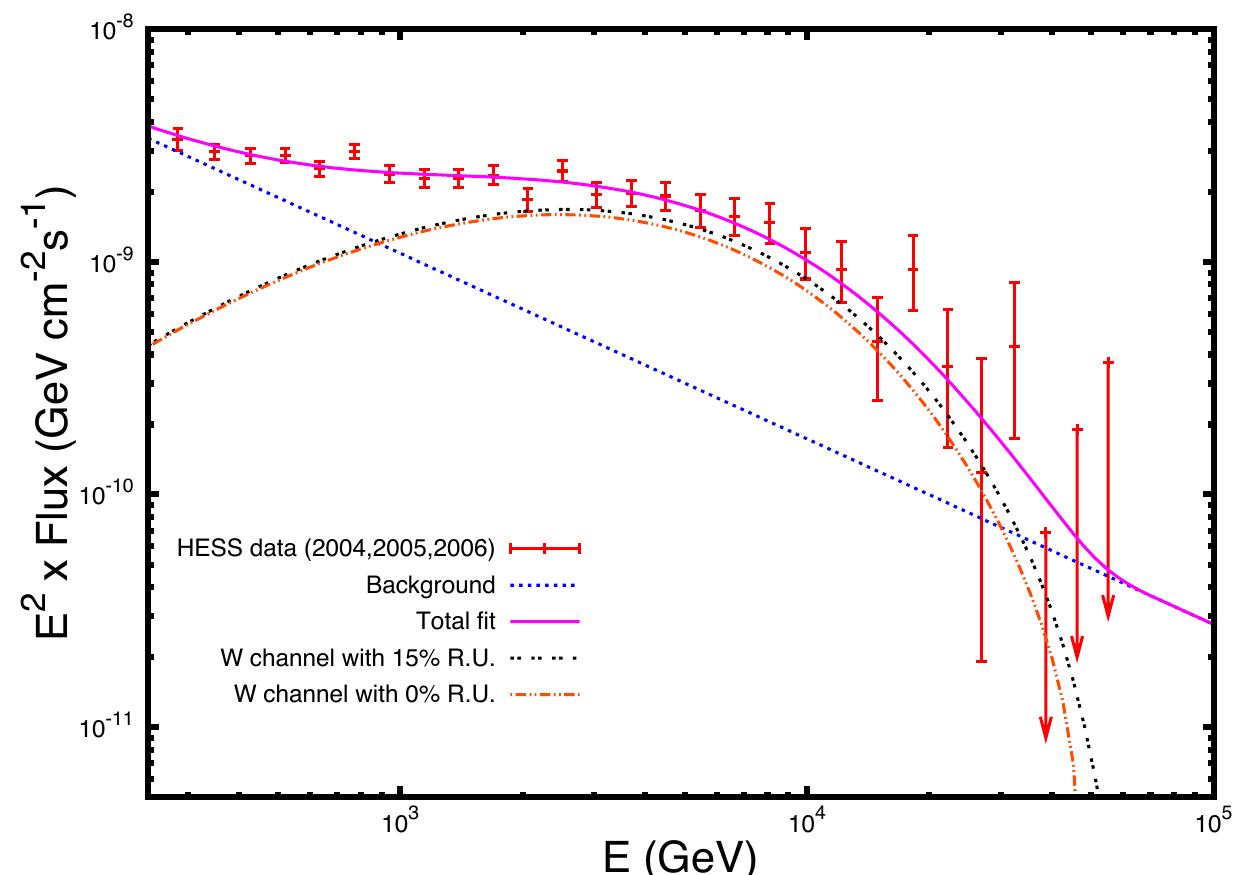}}
\caption {\footnotesize{Same as in Fig. \ref{u fit} but with annihilating DM into $W^+W^-$ gauge bosons}. Electroweak annihilation
channels are softer than hadronic ones. The large uncertainties for data over $\sim 10$ TeV do not allow to discriminate among these
two types of spectra. Both of them are in good agreement with HESS observations.}
\label{W fit}
\end{center}
\end{figure}

\begin{figure}[t]
\begin{center}
\epsfxsize=13cm
\resizebox{7.8cm}{5.6cm}
{\includegraphics{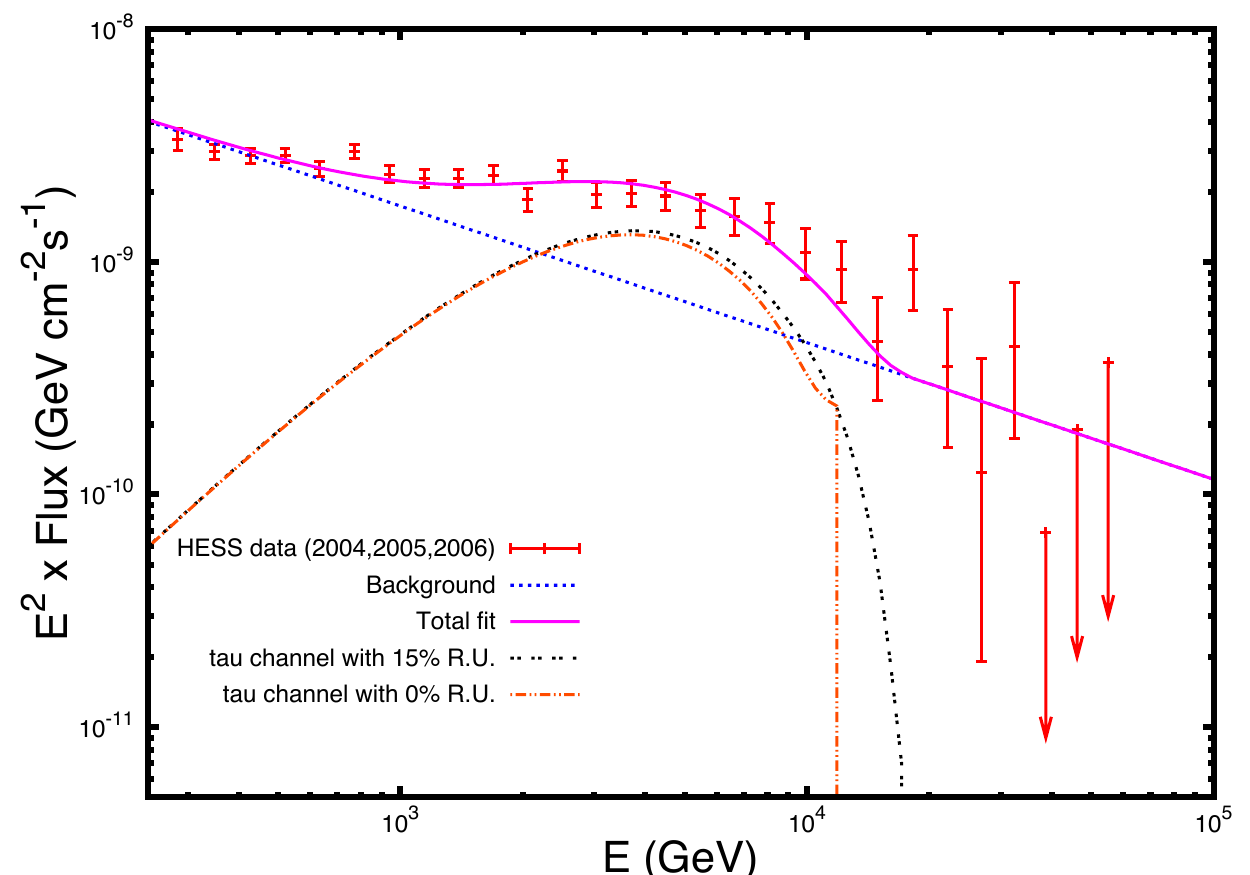}}
\caption {\footnotesize{Same as in Fig. \ref{u fit} but with annihilating DM into $\tau^+\tau^-$.
The poor quality of the fit is evident and common for all the leptonic channels.}}
\label{tau fit}
\end{center}
\end{figure}

\begin{center}
\begin{table*}
\begin{center}
\begin{tabular}{|c|c|c|c|c|c|c|c|}
\hline
\hline
Channel & $M$ (TeV)& $A\,(10^{-7}\,\text{cm}^{-1}\text{s}^{-1/2})$ & $B\,(10^{-4}\, \text{GeV}^{-1/2} \text{cm}^{-1} \text{s}^{-1/2})$
&$\Gamma$ & $\chi^2/\,$dof & $\Delta\chi^2$ &$b$\\
\hline
\hline
$e^+e^-$        & $7.51 \pm 0.11$  	& $8.12 \pm 0.73$ &$2.78\pm0.79$ &$2.55\pm 0.06$ & 2.09& $32.6$ 	&$111 \pm  20$		 \\
\hline
$\mu^+\mu^-$	& $7.89 \pm 0.21$  	&$ 21.2 \pm 1.92$ &$2.81\pm0.53$ &$2.55\pm 0.06$ & 2.04 & $31.4$	&$837 \pm 158$ 	 \\
\hline
$\tau^+\tau^-$	& $12.4 \pm 1.3$  	& $7.78 \pm 0.69$ &$3.17\pm0.62$ &$2.59\pm 0.06$ & 1.59 & $20.6$	&$278 \pm  76$  	 \\
\hline
$u\bar u$       & $27.9 \pm 1.8$  	& $6.51 \pm 0.46$ &$9.52\pm9.47$ &$3.08\pm 0.35$ & 0.78 & $1.2$		&$987\pm 189$ 	 \\
\hline
$d\bar d$       & $42.0 \pm 4.4$  	& $4.88 \pm 0.48$ &$8.26\pm7.86$ &$3.03\pm 0.34$ & 0.73 & $0.0$		&$1257\pm361$	 \\
\hline
$s\bar s$     	& $53.9\pm 6.2$  	& $4.85\pm 0.57$  &$6.59\pm5.43$ &$2.92\pm 0.29$ & 0.90 & $4.1$		&$2045\pm672$ 	 \\
\hline
$c\bar c$       & $31.4\pm 6.0$  	& $6.90 \pm 1.06$ &$53.0\pm157$ &$3.70\pm1.07$ & 1.78& $25.0$ 	          &$1404\pm689$	 \\
\hline
$b\bar b$      	& $82.0 \pm 12.8$  	& $3.69 \pm 0.61$ &$6.27\pm6.07$ &$2.88\pm 0.35$ & 1.32 &$14.2$		&$2739\pm1246$  \\
\hline
$t\bar t$       & $87.7\pm 8.2$  	& $3.68 \pm 0.34$ &$6.07\pm3.34$  &$2.86\pm 0.19$ & 0.88 & $3.6$		&$3116\pm 820$  \\
\hline
$W^+W^-$      	& $48.8 \pm 4.3$  	& $4.98 \pm 0.40$ &$5.18\pm2.23$ &$2.80 \pm 0.15$& 0.84 & $2.6$		&$1767\pm 419$  	 \\
\hline
$ZZ$        	& $54.5 \pm 4.9$  	& $4.73 \pm 0.40$ &$5.38\pm2.45$ &$2.81 \pm 0.16$& 0.85 & $2.9$		&$1988\pm491$  	 \\
\hline
\hline
\end{tabular}
\end{center}
\caption{In this table, the four parameters of the annihilating DM into a single channel fit are presented:
The fitted value of the mass (TeV) of the annihilating WIMPs, normalization factor of the signal $A\,(10^{-7}\,\text{cm}^{-1}\text{s}^{-1/2})$,
normalitazion factor of the gamma ray diffuse emission background $B\,(10^{-4}\, \text{GeV}^{-1/2} \text{cm}^{-1} \text{s}^{-1/2})$,
and spectral index $\Gamma$ of the same background. The $\chi^2$ per degree of freedom (dof), and the value of its
variation $\Delta\chi^2\equiv \chi^2-\chi^2_{d\bar d}$ with respect to the best one ($\chi^2_{d\bar d}$)
is also provided (In the case of four parameters, values of $\Delta \chi^2=4.72\,,9.70$ and $13.3$ correspond
to $68.3\%\,,95.4\%$ and $99.0\%$ confidence level, respectively). Finally, the astrophysical
factor is computed by assuming $\langle \sigma v \rangle = 3\cdot 10^{-26}\; \text{cm}^{3} \text{s}^{-1}$, and presented in units of the the
astrophysical factor associated with a NFW profile: $b\equiv \langle J_{(2)} \rangle/\langle J^{\text{NFW}}_{(2)} \rangle$,
where $\langle J^{\text{NFW}}_{(2)} \rangle\simeq 280 \cdot 10^{23}\; \text{GeV}^2 \text{cm}^{-5}$.}
\label{channels}
\end{table*}
\end{center}

On the other hand, the rest of the computation depends on the nature of the DM particle. If it is meta-stable, photons can
be produced by its decay. In this case ($a=1$): $\zeta^{(1)}_i\equiv\Gamma_i$ is the decay width into SM particles (labeled by the subindex i).
This possibility is not common to all DM candidates. Much more general is the DM annihilation since DM particles are their own antiparticle in
the most part of DM models. The case $a=2$ takes into account the gamma rays coming
from DM annihilation since $\zeta^{(2)}_i\equiv\langle\sigma_i v\rangle$ are the thermal averaged annihilation cross-sections of two
DM particles into SM particles (also labeled by the subindex i). $M$ is the mass of the DM particle, and the number
of photons produced in each annihilating or decaying channel $dN^{(a)}_{i}/dE$, involves decays and/or hadronization of unstable products such
as quarks and leptons. Because of the non-perturbative QCD effects, the
calculation of $dN^{(a)}_{i}/dE$ requires Monte Carlo events generators such
as PYTHIA \cite{pythia}. However, the fact that simulations have to be performed
for fixed DM mass implies that we cannot obtain explicit $M$ dependence for
the photon spectra. In order to overcome this limitation, different
fitting functions have been obtained in \cite{Ce10} for a wide range
of masses which will allow us to include the mass $M$  as an additional
parameter in the fits to HESS data. Those fitting functions are shown to depend
on several parameters whose scaling behaviour with $M$ were also
obtained in \cite{Ce10}. This fact has allowed us to extrapolate the simulated spectra to the
very high energies required in this analysis.

In this letter, we will focus on gamma rays coming from external bremsstrahlung and fragmentation of SM particle-antiparticle pairs produced by
DM annihilation. We will ignore DM decays, the possible production of monoenergetic photons, n-body annhilitations (with $n>2$), or photons produced
from internal bremsstrahlung, that are model dependent. In particular and in order to simplify the discussion and provide useful information for a general analysis, we will consider DM annihilation into each single channel of SM particle-antiparticle pairs, i.e.
\begin{equation}
\label{singlechannel}
\frac{d\Phi_{\text{DM}}}{dE}= A^2 \cdot \frac{dN^{(2)}_{i}}{dE}\,,
\end{equation}
where
\begin{equation}
\label{A}
A^2=\frac{\langle \sigma v \rangle\, \Delta\Omega\, \langle J_{(2)} \rangle_{\Delta\Omega}}{8\pi M^2}
\end{equation}
is a new constant that will be fitted together with the DM particle mass $M$, and the background parameters $B$ and
$\Gamma$. We assume a typical experimental resolution of $15\%$ ($\Delta E/E\simeq0.15$) and a perfect detector efficiency. The results of these 4-parameters fits are summarized on Table \ref{channels}. We can see that the best fit
is provided by the $d\bar d$ channel with $\chi^2/dof=0.74$ for a total of 24 dof. In any case, other hadronic channels
such as $u\bar u$  (see Fig. \ref{u fit}) or $s\bar s$, also provide very good fits within 1$\sigma$. In the same way, softer spectra as the one
provided by $ZZ$, $W^+W^-$  (see Fig. \ref{W fit}) or $t\bar t$ channels are consistent with data without statistical significance difference.
On the contrary, leptonic channels (not only $e^+e^-$, or $\mu^+\mu^-$ but also  $\tau^+\tau^-$,  Fig. \ref{tau fit}), $c\bar c$ and
$b\bar b$ channels are ruled out with more than 99\% confidence level
when compared to the best channel.

It is interesting to note that taking into account all the channels that provide a good fit, the DM mass is constrained to
$15 \; \text{TeV} \lesssim M \lesssim 110 \; \text{TeV}$ within 2$\sigma$.  The lighter values are consistent with
hadronic annihilations ($u\bar u$) and the heavier ones with the annihilation in $t\bar t$, that is more similar to
electroweak channels. On the other hand, at the same 95\% confidence level, the allowed range for the spectral index
of the diffuse background is $2.4 \lesssim \Gamma \lesssim 3.7$. In this case, the lower values are consistent with all
the allowed channels, but the higher values are only accessible to the light quark channels. In any case, as shown in Table I,  all  the channels that provide
a good fit to HESS data are also consistent
with the spectral index observed by Fermi-LAT data: $\Gamma=2.68\pm 0.05$ for energies between
$2 \; \text{GeV} \lesssim E \lesssim 100 \; \text{GeV}$ \cite{ferm} as we have already discussed.

If we know the value of the annihilation cross-section, it is possible to obtain the astrophysical factor. In Table \ref{channels},
$\langle J_{(2)} \rangle$ is given for each channel by assuming
$\langle \sigma v \rangle = 3\cdot 10^{-26}\;\text{cm}^{3} \text{s}^{-1}$
and presented in units of the astrophysical factor associated with a NFW profile:
$b\equiv \langle J_{(2)} \rangle/\langle J^{\text{NFW}}_{(2)} \rangle$.
We see that these results are consistent with $b \sim 10^3$, that are predicted
by simulating the baryonic dissipation effect in the DM halo \cite{Prada:2004pi} as we have commented above.

\begin{table*}
\begin{center}
\begin{tabular}{|c|c|c|c|c|c|c|c|c|}
\hline
\hline
Branons & $M$ (TeV) & $C\; (10^{-2}\, \text{GeV}\, \text{cm}^{-1} \text{s}^{-1/2})$ & $B\; (10^{-4}\, \text{GeV}^{-1/2} \text{cm}^{-1} \text{s}^{-1/2})$
&$\Gamma$ & $\chi^2/\,$dof & $\Delta\chi^2$ &$b$\\
\hline
\hline
W+Z     & $50.6 \pm 4.5$  & $1.57 \pm 0.13$ &$5.27\pm2.32$ &$2.80 \pm 0.15$ & 0.84  &  2.6  & $4843\pm1134$ \\
\hline
\hline
\end{tabular}
\end{center}
\label{DMBra}
\caption{Best fit parameters for branon annihilation. The dominant channels contributing are $W^+W^-$ and $ZZ$.
In the second column $C^2\equiv {\Delta\Omega\, \langle J_{(a)} \rangle_{\Delta\Omega}}/{(8\pi M^2)}$. $b$ is computed
with the model fitted cross-section to WMAP data \cite{WMAP}: $\langle \sigma v \rangle = (1.14\pm0.19)\cdot10^{-26}\; \text{cm}^{3} \text{s}^{-1}$.
Rest of parameters are as in Table I. }
\end{table*}

For a particular DM candidate, several channels can actually contribute. In this situation, the above analysis can be helpful
as a guide but not determinant, since a combination of channels can work in a different way. For example, an interesting DM
candidate which could have high enough mass and account for the right amount of DM in the form of a thermal relic,
is the branon \cite{branons}, associated with brane fluctuations in brane-world models. For masses over $1$ TeV,
the main contribution to the photon spectra comes from branons annihilating into gauge bosons $ZZ$, and $W^+W^-$.
They produce approximately the same amount of $Z$, $W^+$ and $W^-$ since
 $\langle\sigma_{W^+W^-} v\rangle \simeq 2\langle\sigma_{ZZ} v\rangle \simeq M^6/(8\pi^2\, f^8)$,
where $f$ is the brane tension scale (and we are assuming only one branon specie \cite{branons}). As the rest of  channels
can be neglected, and the W and Z channels produce  very similar photon fluxes, the results for the fit with the branon
model lead to very similar results to those obtained by considering both channels individually (see Table II). In this case, as there is a particle model behind, we can deduce the
coupling that leads to the DM abundance consistent with WMAP observations \cite{WMAP}: $f=27.5 \pm 2.4$ TeV, and we can compute the thermal
averaged cross-section: $\langle \sigma v \rangle = \sum_{i=W,Z} \langle \sigma_i v\rangle= (1.14\pm0.19)\cdot10^{-26}\; \text{cm}^{3} \text{s}^{-1}$,
that agrees with the expected order of magnitude.

In this work, we have analyzed the possibility of explaining the gamma ray data observed by HESS from the central part of
the galaxy by being partially produced by DM annihilation. We have proved that even single channel annihilations provide
good fits if the DM signal is complemented with a diffuse background compatible with Fermi LAT observations.
The morphology of the signal is consistent with dark halos compressed by taking into account baryonic dissipation \cite{Blumenthal,Prada:2004pi}.
The DM particle that may have originated these data needs to be heavier than $\sim 10$ TeV.
This makes extremely difficult that
these particles could be observed in direct detection experiments or produced in particle colliders \cite{lab}.
In this sense, the analysis of other cosmic
rays \cite{cosmics}(not only photons, but also antiprotons, positrons, neutrinos,...) from the GC and from other astrophysical objects is fundamental to cross check the hypotheses considered in this work.

\vspace{0.5cm}

{\bf Acknowledgements}
This work has been supported by MICINN (Spain) project numbers FIS 2008-01323, FIS2011-23000, FPA2011-27853-01
and Consolider-Ingenio MULTIDARK CSD2009-00064. JARC acknowledges the kind hospitality of the Institute of
Theoretical Astrophysics at the University of Oslo, Norway.


\end{document}